\documentclass[12pt]{article}
\usepackage{epsfig}
\usepackage{amsfonts}
\begin{document}
\begin{titlepage}
\begin{center}

{\Large Generalized statistical mechanics of cosmic rays}

\vspace{2.cm} {\bf Christian Beck}

\vspace{2.cm}

School of Mathematical Sciences, Queen Mary, University of London,
Mile End Road, London E1 4NS, UK.

\vspace{5cm}

\end{center}

\abstract{We consider a generalized statistical mechanics model
for the creation process of cosmic rays which takes into account
local temperature fluctuations. This model yields Tsallis
statistics for the cosmic ray spectrum. It predicts an entropic
index $q$ given by $q=11/9$ at largest energies (equivalent to a
spectral index of $\alpha =5/2$), and an effective temperature
given by $\frac{5}{9}T_H$, where $kT_H\approx 180$ MeV is the Hagedorn
temperature measured in collider experiments. Our theoretically
obtained formula is in very good agreement with the experimentally
measured energy spectrum of primary cosmic rays.}

\vspace{1.3cm}

\end{titlepage}

More general versions of statistical mechanics, as introduced by
Tsallis \cite{tsa1} and further developed by many others
\cite{abe,tsa2}, have recently been successfully applied to a
variety of complex physical systems. The idea is to maximize more
general entropy measures than the Shannon entropy, which depend on
a parameter $q$ and which lead to generalized versions of
statistical mechanics. Ordinary statistical mechanics is contained
as a special case for $q=1$. Interesting recent physical
applications of the so-called nonextensive formalism obtained for
$q\not=1$ include the statistics of fully developed hydrodynamic
turbulence \cite{BLS}, defect turbulence \cite{daniels},
scattering processes in $e^+e^-$ annihilation \cite{e+-1,e+-2},
heavy ion collisions \cite{ion}, hadron
collisions \cite{para}, and models of vacuum fluctuations
at the Planck scale \cite{book}. There is growing evidence that
the nonextensive formalism, though possessing the mathematical
structure of an equilibrium formalism, is physically often
relevant for nonequilibrium systems with a stationary state that
possess strong fluctuations of an intensive parameter
\cite{wilk,PRL,eddie}.

Very recently Tsallis and Borges \cite{cos1} analysed the
experimentally measured energy spectrum of cosmic rays from a
nonextensive point of view. They showed that a good fit of the
measured spectrum can be obtained by assuming the interplay of two
different nonextensive canonical distributions with two different
entropic indices $q_1$ and $q_2$ and two different inverse
temperatures $\beta_1$ and $\beta_2$. Moreover, Kaniadakis
\cite{cos2} showed that yet another type of generalized
statistical mechanics (based on entropies that are
neither Shannon nor Tsallis) also yields a good fit of the
measured spectrum, using three fitting parameters. Both these
papers represent new interesting ideas, but at the same time they
use various fitting parameters which are not predicted from first
principles. The question arises whether one can construct a
generalized statistical mechanics model of cosmic rays that
predicts the relevant parameters from first principles and that at
the same time well reproduces the measured cosmic ray energy
spectrum using these parameters.

In this letter we consider such a generalized statistical
mechanics model. The model is quite generally of relevance for
particles that are created by scattering processes at very large
energies (as cosmic ray particles are). The basic idea is that at
very high energies the effective thermodynamic interaction volume
is small and hence one expects that there are strong local
temperature fluctuations. A relevant thermodynamic theory of
high-energy scattering processes is the Hagedorn theory
\cite{hage}, which we will generalize here in the sense that we
take into account local temperature fluctuations. Under reasonable
assumptions on the form of the temperature fluctuations this model
predicts Tsallis statistics for the cosmic ray spectrum. More
importantly, it yields a concrete prediction of the entropic index
$q$ (or the differential spectral index $\alpha$ of the cosmic
flux) and of the relevant effective temperature that enters into
the generalized canonical distributions. Our model turns out to
yield a very good fit of the experimentally measured cosmic ray
spectrum. The advantage of our model is that the most relevant
parameters are calculated, not fitted. The more subtle features of
the cosmic ray spectrum (such as the well-known `knee' and
`ankle') are then finally related to the existence of two
different populations of cosmic rays, each being described by the
above nonextensive theory (with the same parameters) but having a
different flux rate.

To start with, let us first have a look at the experimental data
as collected from various experimental groups \cite{chicago}. These
data are displayed in Fig.~1. Also shown is a curve that
corresponds to a prediction of nonextensive statistical mechanics.
Up to energies of $10^{16}$ eV, the measured flux rate of cosmic
ray particles with a given energy is well fitted by a generalized
canonical distribution of the form
\begin{equation}
p(E)=C \cdot \frac{E^2}{(1+\tilde{\beta}(q-1)E)^{1/(q-1)}}.
\label{can}
\end{equation}
Here $E$ is the energy of the particles
\begin{equation}
E=\sqrt{c^2p_x^2+c^2p_y^2+c^2p_z^2+m^2c^4},
\end{equation}
$\tilde{\beta}=(k\tilde{T})^{-1}$ is an effective inverse temperature variable, and $C$ is a
constant representing the total flux rate. For relativistic
particles the rest mass $m$ can be neglected and one has
\begin{equation}
E\approx c |\vec{p}|.
\end{equation}
The distribution (\ref{can}) is a $q$-generalized relativistic
Maxwell-Boltzmann distribution in the formalism of nonextensive
statistical mechanics. These kind of distributions can be directly
obtained by maximizing the Tsallis entropies \cite{tsa1}
\begin{equation}
S_q=\frac{1}{q-1} (1-\sum_i p_i^q )
\end{equation}
and multiplying with the available phase space volume
\cite{e+-1,e+-2,cos1}. The $p_i$ are the probabilities of the
microstates $i$. As seen in Fig.~1, the cosmic ray spectrum is
very well fitted by the distribution (\ref{can}) if the entropic
index is chosen as
\begin{equation}
q=1.215
\end{equation}
and if the effective temperature parameter is given by
\begin{equation}
k\tilde{T}=\tilde{\beta}^{-1}=107 \; \mbox{MeV}.
\end{equation}
Our main aim in the following is to derive the form of the
distribution (\ref{can}) from physically reasonable assumptions,
based on ordinary statistical mechanics but leading effectively to
a nonextensive description. Moreover, we will derive concrete
numerical values for the two parameters $q$ and $\tilde{\beta}$,
thus proceeding from a mere fit towards a theory.

Cosmic ray particles are created in high energy collision
processes as induced by strong galactic and extragalactic
`accelerators' (see e.g.\ \cite{pdg} for a review).
These `accelerators' are astrophysical sources
and may produce a center of mass energy $E_{CMS}$
of the collision process in the GeV and TeV region, but much
larger energies are possible as well. In general, from a
statistical mechanics point of view it is well known that the
largest average temperature that can be reached in  high energy
collision processes is the Hagedorn temperature $T_H$ \cite{hage},
experimentally measured to be about $180\pm 30$ MeV \cite{hagexp1,hagexp2}.
The
Hagedorn temperature is much smaller than $E_{CMS}$ and represents
a kind of `boiling temperature' of nuclear matter at the
confinement phase transition. Even largest $E_{CMS}$ cannot
produce a larger average temperature than $T_H$ due to the fact
that the number of possible particle states grows exponentially.


Let us now provide a plausible physical argument
how nonextensive behavior
comes into play at very large energies. This argument is just
based on ordinary statistical mechanics but takes into
account fluctuations. The
larger the center of mass energy of the collision
process, the smaller the volume probed, due to the uncertainty relation.
This means, the effective interaction volume where a thermodynamic
description of the collision process makes sense will become
smaller and smaller with increasing $E_{CMS}$. However, a smaller
volume means larger temperature fluctuations. It thus makes sense
to consider a generalized Hagedorn theory which takes into
account local temperature fluctuations.

Assume that locally
some value of the fluctuating inverse temperature $\beta$
is given. We then expect
the momentum of a randomly picked particle in this region to be
distributed according to the relativistic Maxwell-Boltzmann
distribution
\begin{equation}
p(E|\beta)=\frac{1}{Z(\beta)}E^2 e^{-\beta E}. \label{max}
\end{equation}
Here $p(E|\beta)$ denotes the conditional probability of $E$ given
some value of $\beta$. We neglect the rest mass $m$ so that
$E=c|\vec{p}|$. The normalization constant is given by
\begin{equation}
Z(\beta)=\int_0^\infty E^2 e^{-\beta E} dE=\frac{2}{\beta^3} .
\end{equation}
Now let us take into account local temperature fluctuations in the
small interaction volumes where cosmic ray particles are produced.
We have to consider some suitable probability density $f(\beta)$
of the inverse temperature in the various interaction volumes.
While the most natural probability distribution of a continuous
random variable taking positive and negative values is the
Gaussian distribution, the most natural distribution of a
continuous random variable living on a positive support (such as
$\beta$) is the $\chi^2$-distribution of degree $n$, i.e.\ the
probability density of $\beta$ is given by
\begin{equation}
f (\beta) = \frac{1}{\Gamma \left( \frac{n}{2} \right)} \left\{
\frac{n}{2\beta_0}\right\}^{\frac{n}{2}} \beta^{\frac{n}{2}-1}
\exp\left\{-\frac{n\beta}{2\beta_0} \right\}. \label{fluc}
\end{equation}
The $\chi^2$-distribution is a typical distribution that naturally
arises in many circumstances, for example if $n$
independent Gaussian random variables $X_i,\;i=1,\ldots ,n$ with
average $0$ are squared and added. If we write
\begin{equation}
\beta:=\sum_{i=1}^{n} X_i^2 \label{Gauss}
\end{equation}
then $\beta$ has the probability density function (\ref{fluc}).
The average of the fluctuating $\beta$ is given by
\begin{equation}
\langle \beta \rangle =n\langle X_i^2\rangle=\int_0^\infty\beta f(\beta) d\beta= \beta_0
\end{equation}
and the variance by
\begin{equation}
\langle \beta^2 \rangle -\beta_0^2= \frac{2}{n} \beta_0^2.
\end{equation}

The observed cosmic ray distribution at the earth does not contain
any information on the local temperature at which the various
particles were produced. Hence we have to average over all
possible fluctuating temperatures, obtaining the measured energy
spectrum as the marginal distribution
\begin{equation}
p(E)=\int_0^\infty p(E|\beta)f(\beta)d\beta . \label{9}
\end{equation}
The integral (\ref{9}) with $f(\beta)$ given
by (\ref{fluc}) and $p(E|\beta)$ given by (\ref{max})
is easily evaluated and one obtains
\begin{equation}
p(E) \sim \frac{E^2}{( 1+\tilde{\beta}(q-1)E)^{\frac{1}{q-1}}}
\label{14}
\end{equation}
with
\begin{equation}
q=1+\frac{2}{n+6} \label{qwert}
\end{equation}
and
\begin{equation}
\tilde{\beta}=\frac{\beta_0}{4-3q}.
\end{equation}
Eq.~(\ref{14}) is just the fitting function used in Fig.~1.

The variables $X_i$ describe the independent degrees of freedom
contributing to the fluctuating temperature. At very large center
of mass energies, the interaction region is very small, and all
relevant degrees of freedom are basically represented by the 3
spatial dimensions into which heat can flow. We may physically
interpret $X_i^2$ as the heat loss in the spatial $i$-direction,
$i=x,y,z$, during the collision process that generates the cosmic
ray particle. The more heat is lost, the smaller is the local
$kT$, i.e. the larger is the local $\beta$ given by (\ref{Gauss}).
The 3 spatial degrees of freedom yield $n=3$ or, according to
(\ref{qwert}),
\begin{equation}
q=\frac{11}{9}=1.222. \label{qmax}
\end{equation}
For cosmic rays $E_{CMS}$ is very large, hence we expect a
$q$-value that is close to this asymptotic value. The fit in
Fig.~1 in fact uses $q=1.215$, which agrees with the predicted
value in eq.~(\ref{qmax}) to about 3 digits. It also coincides
well with the fitting value $q=1.225$ used by Tsallis et al.
\cite{cos1} using multi-parameter generalizations of nonextensive
canonical distributions. 

For smaller center of mass energies, the
interaction region will be bigger and more effective degrees of
freedom within this bigger interaction region can contribute to
the fluctuating temperature. Hence we expect that for finite
$E_{CMS}$ $n$ will be a bit larger than 3, or $q$ will be slightly
smaller than $11/9$. A good parametrization, in agreement with
experimentally measured cross sections in $e^+e^-$ annihilation,
is \cite{e+-2}
\begin{equation}
q(E_{CMS})= \frac{11-e^{-E_{CMS}/E_0}}{9+e^{-E_{CMS}/E_0}}
\end{equation}
with $E_0\approx 45.6$ GeV.

Let us now predict a value for the other `fitting parameter'
entering into the nonextensive canonical distribution, the
effective temperature $k\tilde{T}\approx 107$ MeV. The Hagedorn
temperature is experimentally measured to 
be about $kT_H \approx 180$ MeV \cite{hagexp1,hagexp2}, which is the right
order of magnitude. But why is the effective temperature fitting
the cosmic ray spectrum smaller by a factor of about 0.6 as
compared to the Hagedorn temperature?

A possible answer could be related to the way the Hagedorn
temperature $T_H$ is experimentally determined. The experimental
measurements are evaluated under the assumption that ordinary
relativistic Maxwell-Boltzmann (MB) statistics applies, which is a
good approximation for relatively small $E_{CMS}$. Assuming MB
statistics
\begin{equation}
p(E)=CE^2e^{-\beta E},
\end{equation}
the temperature $\beta^{-1}$ can be estimated from the maximum
$E^*$ of the distribution as
\begin{equation}
kT^{MB}=\frac{1}{2}E^*.
\end{equation}
This relation immediately follows from evaluating $p'(E^*)=0$.
However, suppose that in reality at sufficiently large $E_{CMS}$
Tsallis statistics
\begin{equation}
p(E)=C \frac{E^2}{(1+(q-1)\tilde{\beta} E)^{\frac{1}{q-1}}}
\end{equation}
applies, due to the fluctuations
mentioned before. Then the relation between the maximum and the temperature
changes. One obtains from evaluating $p'(E^*)=0$ the relation
\begin{equation}
\tilde{\beta}^{-1}=k\tilde{T}=\frac{3-2q}{2} E^*=(3-2q)kT^{MB}.
\label{tbeta}
\end{equation}
In collider experiments, the value of the maximum $E^*$ is seen to be
essentially independent of $E_{CMS}$ \cite{e+-2}.
An experimentally determined Hagedorn temperature (under the
assumption of MB statistics) of $180\pm 30$ MeV and a value of $q=1.215$
thus yields the effective temperature $k\tilde{T}=0.57 \cdot
kT^{MB}= (103 \pm 17)$ MeV. This just coincides (within the error bars)
with the observed
effective temperature that fits the data in Fig.~1. It 
also coincides
with one of the two temperature fitting parameters (96.15 MeV)
used by Tsallis et al. in
\cite{cos1}. For
$E_{CMS}\to \infty$ one obtains from eq.~(\ref{qmax}) and
(\ref{tbeta}) the relation
\begin{equation}
k\tilde{T}= \frac{5}{9}kT^{MB}.
\end{equation}

In this way all relevant parameters determining the shape of the
cosmic ray energy spectrum are predicted from a simple theoretical
consideration (except of course for the total flux rate of the
cosmic rays, represented by the constant $C$).

To explain the `knee' and `ankle' in the cosmic ray spectrum we
adopt the standard view \cite{pdg}. It is reasonable to assume
that the knee at $E\approx 10^{16}$ eV is simply produced by the
fact that one has reached the maximum energy scale to which
typical galactic accelerators can accelerate. This naturally
implies a rapid fall in the number of observed events with a
higher energy, i.e. a steeper slope in Fig.~1 between about
$10^{16}$ and $10^{19}$ eV. The `ankle' at $E\approx 10^{19}$ eV
is then produced by the fact that a higher energy population of
cosmic ray particles takes over from a lower energy population. 
This higher energy
population may have 
a different origin (for example,
extragalactic origin). The new population has a much smaller flux
rate but can reach much larger energies. Naturally, the cosmic
accelerators underlying the production process of this new species of
cosmic rays must have a much larger center of mass energy
$E_{CMS}$ than the ankle energy $\sim 10^{19}$ eV, so $q$ should
be quite precisely given by its asymptotic value $11/9$, whereas
the effective temperature $T$ should be the same as before. The
dashed line in Fig.~1 corresponds to our formula with $q=11/9$,
$k\tilde{T}=107$ MeV and a flux rate that is smaller by a factor $1/50$ as
compared to the high-flux generation of cosmic rays. It is
consistent with the data.

Our nonextensive theory predicts power-law behavior of the
measured energy spectrum. For large $E$ one has $p(E)\sim
E^{-\alpha}$ where the differential spectral index $\alpha$ is given by
\begin{equation}
\alpha =\frac{1}{q-1}-2.
\end{equation}
$q=1.215$ implies $\alpha =2.65$ (for moderately large energies),
whereas for largest energies the asymptotic value $q=11/9$ implies
$\alpha =5/2$. As shown in Fig.~2, the largest-energy events are
compatible with such an asymptotic power law exponent, though of
course the error bars of the experimental data are huge.

To conclude, our simple Hagedorn model based on ordinary
statistical mechanics with local fluctuations of temperature very
well reproduces the experimentally measured cosmic ray spectrum.
By averaging over the temperature fluctuations one effectively
ends up with a nonextensive theory, which describes the statistics
of cosmic ray particles in a generalized statistical mechanics
setting. Our model allows for concrete predictions of the relevant
entropic index $q$ and of the relevant effective temperature
parameter $\tilde{T}$, and these predicted numerical values are in good
agreement with the measured data.

\newpage

\epsfig{file=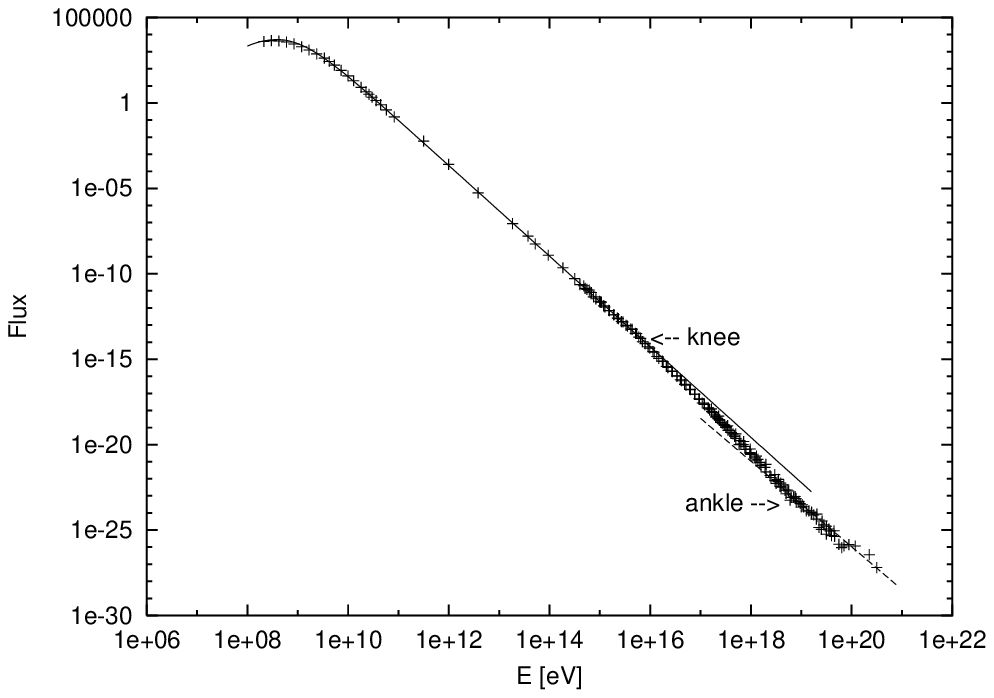}

\vspace{2cm}

{\bf Fig.~1} Measured energy spectrum of primary cosmic rays (in
units of $m^{-3}s^{-1}sr^{-1} GeV^{-1}$)
as listed in \cite{chicago}. The solid line is the
prediction (\ref{can}) with $q=1.215$, $k\tilde{T}=107$ MeV and $C=5\cdot
10^{-13}$ in the above units. The dashed line is eq.~(\ref{can}) with $q=11/9$,
$k\tilde{T}=107$ MeV and $C$ smaller by a factor $1/50$.

\newpage

\epsfig{file=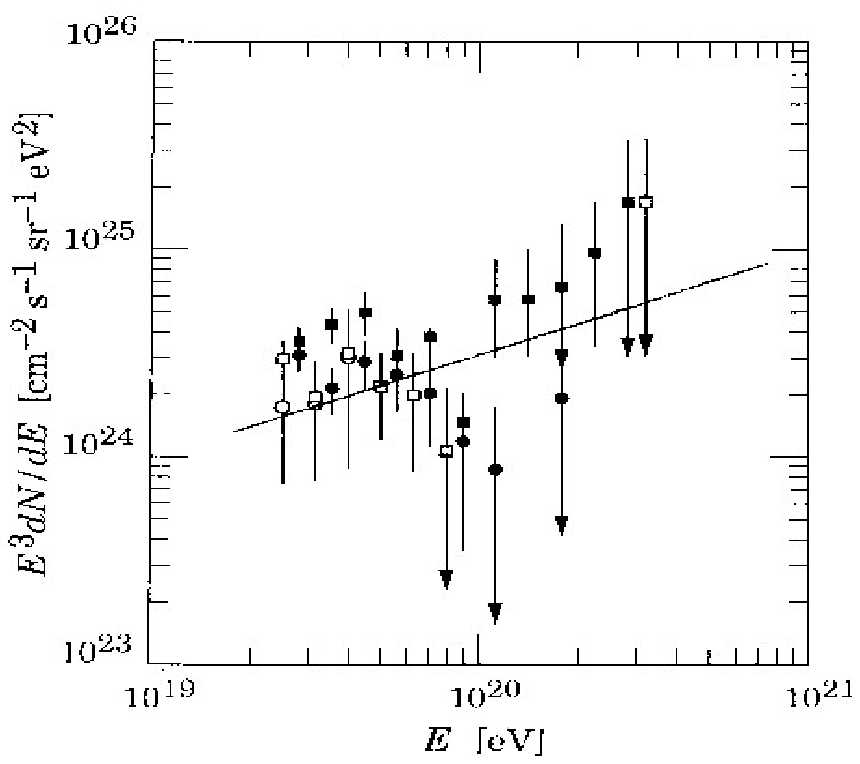}

\vspace{1cm}

{\bf Fig.~2} Measured cosmic ray energy spectrum $E^3 \cdot dN/dE$
at largest
energies (data from \cite{pdg,high1,high2,high3}). The straight line is
the power law prediction with exponent $\alpha =5/2$ (corresponding
to $q=11/9$).

\end{document}